\documentclass{article}
\usepackage{epsfig,amsmath,amssymb,enumitem,authblk,url}
\usepackage[papersize={8.5in,11in}]{geometry}
\usepackage{graphicx}
\newcommand{\be}{\begin{equation}}
\newcommand{\ee}{\end{equation}}
\begin{document}
\title{Implications of the Milky Way Declining Rotation Curve}
\author[1,2]{J. W. Moffat}
\author[1]{H. Sharron}
\author[1]{V. T. Toth}
\affil[1]{\textit{Perimeter Institute for Theoretical Physics, Waterloo, Ontario N2L 2Y5, Canada}}
\affil[2]{\textit{Department of Physics and Astronomy, University of Waterloo, Waterloo,
Ontario N2L 3G1, Canada}}
\maketitle




\begin{abstract}
Almost all spiral galaxies have been observed to have flattening rotation curves. The new Gaia DR3 released data shows a Milky Way sharply Keplerian declining rotation curve, starting at $\sim 16$ kpc and ending at 26.5 kpc. The data reduces the total Milky Way mass by an order of magnitude, $M=2.06\times 10^{11}M_{\odot}$, compared to the standard required dark matter halo mass, $(2-5)\times10^{12}M_\odot$. Newtonian and modified gravity (MOG) fits are applied to the Gaia DR3 rotation curve data. The fit obtained using MOG has a total mass of $M\sim1.3\times 10^{11}M_{\odot}$, while the Newtonian fit predicts a mass of $M\sim2\times 10^{11}M_{\odot}$. These are in excess of the estimated visible baryon mass of the Milky Way, $M_b\sim (0.6-1.0)\times 10^{11}M_{\odot}$. It is possible that if the cicumgalactic (CGM) plasma-gas continues to be confirmed experimentally, then the additional baryon mass required to account for the estimated total Milky Way mass could be attributed to the CGM hot plasma-gas halo. 
\end{abstract}

\section{Introduction}

Extended flat rotation curves of spiral galaxies are considered as evidence of the presence of dark matter in the form of extended halos of dark matter surrounding them, or the need for a modified theory of gravity. Because our location in the Milky Way prevented us from a straightforward determination of the rotation curve, early rotation curve investigations of rotation curves were made beyond our own galaxy. 

Before the early 1960s, the outer Milky Way rotation curve was considered to be Keplerian, but evidence since then has shown that its outer curve is flat, albeit with significant uncertainties. However, the rotation curve determined by Sofue et al.~\cite{Sofue} is consistent with a decreasing rotation curve from $\sim 16$ to $23$ kpc, but with huge uncertainties~\cite{MoffatTothMW}. 

Significant progress in determining the Milky Way rotation curve came with the European Space Agency's Global Astrometric Interferometer for Astrophysics (Gaia) mission\footnote{\url{https://www.esa.int/Science_Exploration/Space_Science/Gaia}}, whose proper motion measurements have allowed 3D velocity measurements. The Gaia data release 2 (DR2) measurements produced a new accurate rotation curve~\cite{Eilers,Mroz}. The Gaia DR3 measurements~\cite{Jiao,Labibi,Zhou,Wang,Ou} have provided improved parallaxes and proper motions with increased accuracy, up to twice the galactocentric radius compared to DR2. The results show a sharply decreasing rotation curve for the Milky Way. The estimated total mass of the Milky Way is significantly lower, $M=2.06_{-0.13}^{+0.24}\times 10^{11} M_{\odot}$, compared to the previously required dark matter halo mass $M\sim (2-5)\times 10^{12}M_{\odot}$~\cite{Boylan-Kolchin,Kafle}.

It is possible that the difference in methodology contributes to the observed discrepancy between the Milky Way's sharply declining rotation curve and the mainly flat rotation curves observed in external galaxies. The external galaxy rotation curves are based on the gas and dust components, often derived from observations of neutral hydrogen gas (HI) using radio telescopes or from optical emission lines and Doppler shift measurements. These methods provide an averaged rotation curve, but lack the ability to resolve individual stellar motion. Being inside the Milky Way allows us to measure three-dimensional motions of individual stars, offering a uniquely accurate and detailed rotation curve not possible for external galaxies. 

MOG is a modified theory of gravitation that is characterized by a length or scale dependence. The weak field MOG  gravitational acceleration law predicts that Newtonian gravity dominates in the solar system system~\cite{Tremaine2024} and also in wide binaries~\cite{Moffatwidebinaries}. 

We investigate the prediction of modified gravity MOG~\cite{Moffat2006,Moffat2020} for the significant decline of the Milky Way rotation curve, showing that the prediction can fit the Gaia DR3 data, based on a baryon source model.
We find that the fits to the Milky Way Gaia DR3 data for the rotation curve need a baryon mass $M_b\sim 1.3\times 10^{11} M_\odot$, which represents only a modest increase over the estimated {\em visible} stellar and gas baryon mass of $(0.6 - 1)\times 10^{11}M_{\odot}$~\cite{Nicastro1}, and may be consistent with the concept of missing baryons.

In cosmology, the missing baryon problem is an observed discrepancy between the detected baryon mass amount shortly after the Big Bang and that detected in more recent epochs. The Big Bang nucleosnythesis constrains the abundance of baryons in the early universe to approximately $4.8\%$ of the matter-energy content of the universe. A census of baryons in the recent universe accounts for less than half of that amount~\cite{Shull,Peebles}. 

The missing baryon-mass
problem is present at all halo scales, from dwarfs to elliptical galaxies. It includes galaxy groups and clusters~\cite{Nicastro1,McGaugh,Bregman,Nicastro2,Zhang,
Odorico,Macquart,Tumlinson,Anderson}. Galaxy disks and halos only contain $\sim 20\%$ of the expected baryons. Large volumes of space surrounding the stellar disks are known to host clouds of hot gas, $T\sim 10^6-10^7$ K, bound by gravitation to the galaxy as a massive baryon envelope. The estimated circumgalactic masses (CGM) of X-ray halos within virial radius correspond to a mass $M_{CGM}\simeq (1-1.7)\times 10^{11}(Z/0.3Z_{\odot})^{-1}M_{\odot}$, where $0.3Z_{\odot}=\langle Z\rangle$ is the average metallicity of the gas. This corresponds to closing the galaxy baryon mass $M_{CGM}/M_{\rm missing}\sim (1-1.7)(Z/0.3Z_{\odot})$. 

The standard dark matter galaxy halos require a dark matter halo with masses up to 
$M_{\rm halo}\sim 5\times 10^{12} M_{\odot}$. This is true for the Milky Way and this dark matter mass is used to fit the $\Lambda$CDM model~\cite{Blanchard}.

Such a massive halo does not appear to be consistent with the Gaia observations, nor is it consistent with certain families of modified gravity theories. In particular, in ref.~\cite{Blanchard}, it is demonstrated that the modified gravity MOND model~\cite{McGaugh,Milgrom} cannot fit the Gaia DR3 rotation curve and fit other galaxy rotation curves. The MOND prediction for the rotation curve is excluded by $5\sigma$.

In addition to investigating MOG, we also obtain a fit to the rotation curve using Newtonian gravity. The fit demands a larger total mass for the Milky Way, $M=M_b+M_{\rm dm}$, given by $M\sim2\times 10^{11}M_{\odot}$. Nonetheless, the dark matter mass $M_{\rm dm}$ is still significantly reduced from the standard dark matter halo mass, $M\sim (2-5)\times 10^{12}M_{\odot}$~\cite{Boylan-Kolchin,Kafle}. 

\section{The MOG Field Equations}

MOG, also known by the acronym STVG for Scalar--Tensor--Vector Gravity, is a theory of gravitation that involves, in addition to the metric field of general relativity, a vector field and scalar degrees of freedom. The core phenomenology of MOG is that while the tensor field couples to matter more strongly than in general relativity, this coupling is partially canceled by a vector field and the resulting repulsion. The vector field is not massless, therefore its range is finite; gravity, therefore, regains its full strength outside the range of the vector force, on galactic and extragalactic scales.

The MOG theory is usually introduced in the form of an action principle and the resulting field equations~\cite{Moffat2006}. Here we present a brief overview of key equations and the nonrelativistic, weak field acceleration law.

The gravitational coupling strength $G$ is defined by $G=G_N(1+\alpha)$, where $G_N$ is the Newtonian gravitational constant and $\alpha$ is a scalar degree of freedom that measures the deviation from General Relativity~\cite{Moffat2006,Moffat2020}. The field equations are given by (we use the metric signature $(+,-,-,-)$ and units with $c=1$):
\be
\label{Gequation}
G_{\mu\nu}=8\pi GT_{\mu\nu},
\ee
\be
\label{Bequation}
\nabla_\nu B^{\mu\nu}=\frac{1}{\sqrt{-g}}\partial_\nu(\sqrt{-g}B^{\mu\nu})=J_M^\mu,
\ee
where $G_{\mu\nu}=R_{\mu\nu}-\frac{1}{2}g_{\mu\nu}R$, 
$\nabla_\mu$ denotes the covariant derivative with respect to the metric $g_{\mu\nu}$, $g={\rm det}(g_{\mu\nu})$ and $B_{\mu\nu}=\partial_\mu\phi_\nu-\partial_\nu\phi_\mu$.  Moreover, $J_M^\mu=\kappa\rho u^\mu$, $\rho$ is the density of matter and field energy, 
$\kappa=\sqrt{\alpha G_N}$ and $u^\mu=dx^\mu/ds$. The energy-momentum tensor is
\be
T_{\mu\nu}=T^M_{\mu\nu}+T^\phi_{\mu\nu},
\ee
where
\be
T^\phi_{\mu\nu}=-\biggl({B_\mu}^\alpha B_{\alpha\nu}-\frac{1}{4}g_{\mu\nu}B^{\alpha\beta}B_{\alpha\beta}\biggr).
\ee
The conservation equations are given by
\be
\label{conservequation}
\nabla_\nu(T^{\mu\nu}_M+T^{\mu\nu}_\phi)=0.
\ee

The gravitational coupling strength $G=G_N(1+\alpha)$ scales 
with $\alpha$, depending on the strength of the gravitational field. A key premise of MOG is that all baryonic matter possesses, in proportion to its mass $M$, positive gravitational charge $Q_g=\kappa M$. This gravitational charge serves as the source of the vector field 
$\phi_\mu$. MOG fits to galaxy rotation curves, galaxy clusters and cosmology without dark matter have been published~\cite{BrownsteinMoffat2007,MoffatRahvar2013,MoffatRahvar2014,
GreenMoffat2019,DavariRahvar2020,IsraelMoffat2018,MoffatHaghighi2017,
MoffatTothCMB}.

The modified Newtonian acceleration law for weak gravitational fields and for a point particle can be written
as~\cite{Moffat2006}:
\be
\label{MOGacceleration}
a_{\rm MOG}(r)=-\frac{G_NM}{r^2}[1+\alpha-\alpha\exp(-\mu r)(1+\mu r)].
\ee
This reduces to Newton's gravitational acceleration in the limit
$\mu r\ll 1$. In the limit that $r\rightarrow\infty$, we get,
for approximately constant $\alpha$ and $\mu$:
\be
\label{AsymptoticMOG}
a_{\rm MOG}(r)\approx -\frac{G_N(1+\alpha)M}{r^2}.
\ee
The circular rotational velocity is calculated from the formula:
\be
v_c(r)=(ra_{MOG}(r))^{1/2}=\bigg\{\frac{G_NM(r)}{r}[1+\alpha-\alpha
\exp(-r/r_0)(1+r/r_0)]\bigg\}^{1/2}.
\ee
where $\mu=1/r_0$.

The ratio of the MOG acceleration and the Newtonian acceleration is given by
\be
\label{gravityratio}
\frac{a_{\rm MOG}}{a_N}=1+\alpha-\alpha\exp(-r/r_0)(1+r/r_0).
\ee
For a distributed baryonic matter source, the MOG weak field acceleration law becomes~\cite{GreenMoffat2019}:
\be 
a_{\rm MOG}=-G_N\int d^3x^\prime\frac{\rho_b(x^\prime)(x-x^\prime)}{|x-x^\prime|^3}
[1+\alpha-\alpha\exp(-\mu|x-x^\prime|)(1+\mu|x-x^\prime|)],
\ee
where $\rho_b$ is the total baryonic mass density.

\section{The Milky Way rotation curve}

To compute the rotation curve of the Milky Way, we need a model for the distribution of the matter components of the Milky Way. A typical model used is given by Binney and Tremaine~\cite{BinneyTremaine}, Salas et. al.,~\cite{deSalas} and Jiao et al.~\cite{Jiao}. This model consists of a spherical bulge described by a Hernquist potential, an axi-symmetric double exponential thin stellar disk model and multiple gas disks.

In reference~\cite{GreenMoffat2019}, mass models based on a bulge and disk model were derived independently of the SPARC catalogue data of 149 rotationally supported and irregular galaxies. The radial rotation velocity curves were derived using MOG and compared to the Newtonian acceleration law. 

We shall use a simpler spherically symmetric mass model that encompasses all the stellar and gas baryon matter out to a Milky Way radius $R_0$. The mass model is given by
\begin{equation}
        M(r)= \begin{cases}
            M_0 \bigg[\exp{\Big[{1-\Big(\dfrac{R_0}{r}\Big)^{e_1}}\Big]}\bigg]^{e_2} & r < R_0 \\
            M_0 & r \geq R_0
        \end{cases}
\end{equation}
where $M_0$ is the mass for the exterior vacuum MOG acceleration formula (\ref{MOGacceleration}) and $e_1$ and $e_2$ are adjustable parameters. The rotation curve to be fitted to the Gaia DR3 data is far enough away from the central bulge and disk to allow for our spherically symmetric mass model to serve as a good approximation to the Milky Way mass distribution. It can also account for an extended baryon plasma-gas mass halo enveloping the Milky Way.
\begin{figure}[b!]
    \centering
    \includegraphics[width=0.5\linewidth]{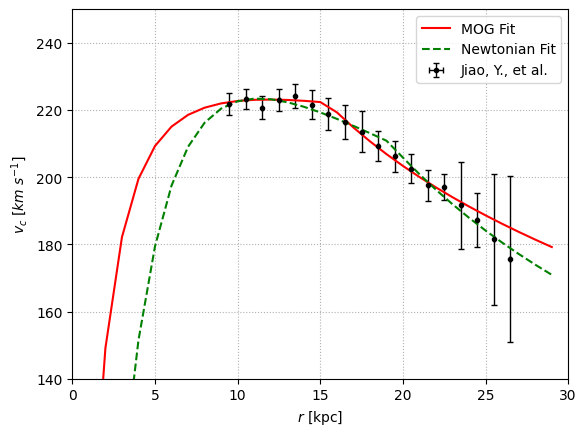}
    \includegraphics[width=0.49\linewidth]{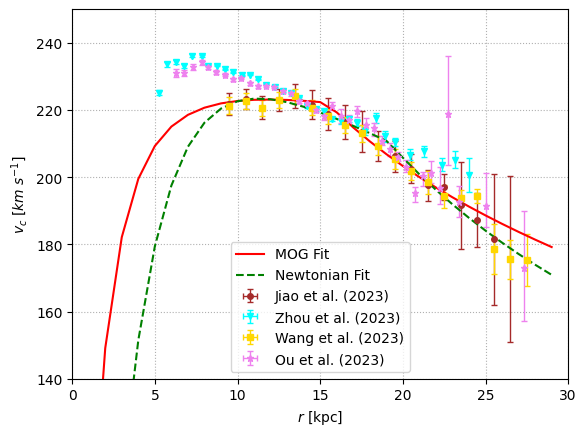}
    \caption{\textbf{Left}: This plot shows two calculated rotation curve fits to the recent Gaia DR3 data shown in black with error bars~\cite{Jiao}. The first curve in red is a MOG fit with parameters $\alpha=0.91$, $\mu=0.0803$  ${\rm kpc}^{-1}$, $M_0=1.34\times10^{11}M_\odot$, $e_1=0.618$ and $e_2=1.018$ and $R_0=15.4$ kpc, with $\chi_\nu^2=0.15$. The second curve shown in dotted green is a Newtonian fit with parameters $\alpha=0$, $M=1.97\times10^{11}M_\odot$, $e_1=1$, $e_2=0.59$, and $R_0=19.1$ kpc, with $\chi_\nu^2=0.19$. \textbf{Right}: The plot shows four rotation curve data sets from Jiao et al., Zhou et al., Wang et al., Ou et al.~\cite{Jiao,Zhou,Wang,Ou}, in circular brown, triangular cyan, square gold and star violet data points respectively. The MOG and Newtonian fits from the left is shown in red and dotted green, respectively.}
    \label{fig:main}
\end{figure}
 \begin{figure}[t!]
     \centering
     \includegraphics[width=0.75\linewidth]{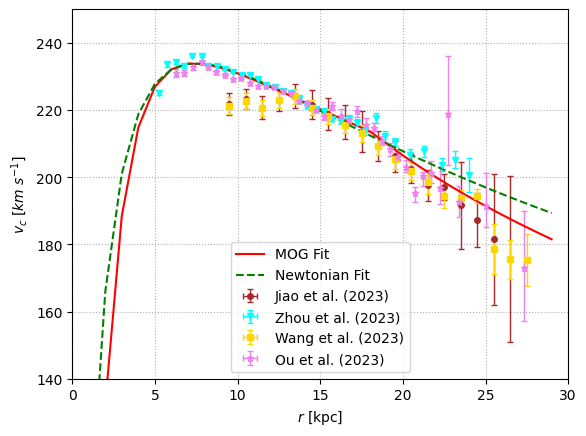}
     \caption{The Gaia DR3 rotation curve data sets Jiao et al., Zhou et al., Wang et al., Ou et al.,~\cite{Jiao,Zhou,Wang,Ou} are shown as circular brown, triangular cyan, square gold and star violet data points respectively. A MOG weighted fit on all four data sets is shown in red, with parameters $M_0=1.28\times10^{11} M_\odot$, $R_0=18.1$ kpc, $e_1=1.107$, $e_2=0.266$,
     $\alpha=0.96$, and $\mu=0.096$ kpc$^{-1}$, with $\chi_\nu^2=4.15$. The weighted Newtonian
     fit to all four data sets in shown in dotted green, with parameters $M_0=2.58\times10^{11} M_\odot$, $R_0=33.8$ kpc, $e_1=0.597$, $e_2=0.683$, with $\chi_\nu^2=4.83$.}
     \label{fig:milkyway}
 \end{figure}
In Figure \ref{fig:main}, we show a fit to the rotation curve Gaia DR3 data~\cite{Jiao}. 
This data set is chosen as it thought to be the most reliable. The latest paper of the four, it considered the rotation curve analysis performed by Zhou et al., Wang et al., Ou et al~\cite{Zhou,Wang,Ou}, and had more reliable systematics. This Figure~\ref{fig:main} shows a MOG best fit (red solid line) with parameters $\alpha=0.91$, $\mu=0.0803~{\rm kpc}^{-1}$, $M_0=1.34\times 10^{11}M_{\odot}$, $R_0=15.4$ kpc, $e_1=0.618$ and $e_2=1.018$, with $\chi_\nu^2=0.15$. The dotted green curve is a Newtonian fit 
with parameters $\alpha=0$, $R_0=18.03$ kpc, $e_1=1$, $e_2=0.59$ and $M_0=1.97\times 10^{11}M_{\odot}$, with $\chi_\nu^2=0.19$. This corresponds to a Newtonian acceleration:
\be
a_N(r)=\frac{G_NM(r)}{r^2}.
\ee
For the MOG red curve fit in Figure \ref{fig:main}, we have a ratio of MOG acceleration to Newtonian acceleration at $r=20$ kpc of $a_{\rm MOG}/a_N=1.43$.

 \begin{table}[b!]
     \caption{The parameters and $\chi_\nu^2$ of the MOG and Newtonian fits performed in 
     Figure \ref{fig:main} and Figure \ref{fig:milkyway} are shown in two columns on the
      left and right respectively.}
     \label{tab:milkyway}
     \centering
     \begin{tabular}{c|c|c|c|c}
          & MOG Fit (Jiao) & Newtonian Fit (Jiao) & MOG Fit (All) & Newtonian Fit (All)\\
          \hline
         $M_0$ & $1.34\times10^{11} M_\odot$ & $1.97\times10^{11} M_\odot$ & $1.28\times10^{11} M_\odot$ & $2.58\times10^{11}M_\odot$\\
         \hline
         $R_0$ & 15.4 kpc & 19.1 kpc & 18.1 kpc & 33.8 kpc \\
         \hline
         $e_1$ & 0.618 & 1 & 1.107 & 0.597\\
         \hline
         $e_2$ & 1.018 & 0.59 & 0.266 & 0.683\\
         \hline
         $\alpha$ & 0.91 & 0 & 0.96 & 0\\
         \hline
         $\mu$ & 0.0803 kpc$^{-1}$ & 0 & 0.096 kpc$^{-1}$ & 0\\
         \hline
         $\chi_\nu^2$ & 0.15 & 0.19 & 4.15 & 4.83 
     \end{tabular}
 \end{table}
In Figure \ref{fig:milkyway}, a MOG and Newtonian best fit weighted to all four data sets is shown in red and dotted green respectively. This fit was performed to better incorporate all of the available data instead of fitting only one data set as in Figure \ref{fig:main} to Jiao et al. The MOG weighted fit has parameters $M_0=1.28\times10^{11} M_\odot$, $R_0=18.1$ kpc, $e_1=1.107$, $e_2=0.266$, $\alpha=0.96$, and $\mu=0.096$ kpc$^{-1}$, with $\chi_\nu^2=4.15$. The weighted Newtonian fit has parameters $M_0=2.58\times10^{11} M_\odot$, $R_0=33.8$ kpc, $e_1=0.597$, $e_2=0.683$, with $\chi_\nu^2=4.83$. The larger $\chi_\nu^2$ values in comparison with the fits shown in Fig.~\ref{fig:main} are a result of the much tighter error bars, especially on the Zhou and Ou data sets, and the deviation of their values from the Jiao and Wang data at lower radii.

The Newtonian estimate of Milky Way mass, $M\sim 2\times 10^{11}M_{\odot}$ can be determined approximately from the measured circular velocity using the relation: 
\be
M=v_c^2R/G_N,
\ee
where $v_c$ is the circular velocity at the radius $R$. Applying this estimate, it has been determined that the enclosed mass varies little beyond $R > 19$ kpc.

The excess mass can be attributed to dark matter. The excess by a factor $\sim 2-3$ can also be attributed to the CGM plasma-gas. If it is attributed to dark matter mass, then this represents a significant reduction in the dark matter factor of order 10-16 required to fit the galaxy dark matter halos of many spiral galaxies. There has been no success up till now to detect dark matter particles. There has been success in detecting the hot CGM plasma-gas in galaxies and in the Milky Way galaxy~\cite{Nicastro1,Shull,Peebles,McGaugh,Bregman,Nicastro2, Zhang,Odorico,Macquart,Tumlinson,Locatelli}. 

There are galaxies that exhibit no or negligible dark matter that are accounted for by MOG. These include the compact galaxy NGC1277~\cite{MoffatTothNGC1277} and the ultra diffuse galaxies DF2 and DF4~\cite{Dokkum,Pina,MoffatTothDF2}. 

The Newtonian fit has a mass of $\sim2\times 10^{11}M_\odot$.  This mass estimate requires reduction in dark matter by an order of magnitude from $\sim(2-5)\times10^{12}M_\odot$ ~\cite{Boylan-Kolchin,Kafle} to $1.97\times10^{11}M_\odot$. Instead of dark matter, this excess mass could be accounted for by a CGM hot gas~\cite{Nicastro2} with mass estimates $M_{\rm CGM}$.

 The results from fitting the rotation curve data have significant implications for our understanding of galaxy dynamics, galaxy formation and evolution, and cosmology. If we attribute the excessive baryons in the Milky Way to the detected hot CGM gas, then we have to understand why the Milky Way is significantly different in its composition of matter and its evolution, and how our understanding of the composition of matter in cosmology has to be investigated. The consequences for dark matter and modified gravity models have to be considered.

\section{Conclusions}

We have investigated the implications of the Gaia DR3 for the rotation curve and the estimated mass of the Milky Way. Applying the modified gravity MOG to fit the rotation curve data, we find that fits need a mass $M\sim 1.3\times 10^{11}M_{\odot}$. The Newtonian-Keplerian decline between $\sim 16$ to 27 kpc is consistent with Newton's shell theorem that most of the mass is located within the the central Milky Way bulge and disk. The mass $M\sim 1.3\times 10^{11}M_{\odot}$ of the MOG fit, is in excess of the estimated {\em visible} baryon mass of the Milky Way $M_b\sim (0.34-1)\times 10^{11}M_{\odot}$. This additional excess mass can be attributed to dark matter or the missing baryon mass in the Milky Way. However, recent investigations have claimed that galaxies have an additional baryon mass, detected as a hot $T\sim 10^6-10^7$ K CGM hot plasma-gas, which may imply that the missing mass is baryonic in nature.

The Newtonian-Keplerian decline of the rotation curve can also be fitted by Newtonian gravity, with a total mass that can include dark matter $M_{\rm dm}$. The amount of dark matter is significantly smaller than the dark matter normally required to fit spiral galaxies enveloped by a dark matter halo, as predicted by the standard $\Lambda$CDM model.

According to the Gaia DR3, the Milky Way rotation curve cannot be consistent with a flat rotation curve at a significance of $3\sigma$. There are spiral galaxies such as UGC 4458 and NGC 2599, showing a Keplerian decline of their rotation curves. The Milky Way may be exceptional as a spiral galaxy showing a Keplerian decline beginning $\sim16$ kpc in its rotation curve. This exception may be due to its relative quiet past history~\cite{Hammer,Haywood}, having experienced no major merger for $\sim 10$ Gyr. 

The Gaia DR3 methodology concentrated on precisely measured
stellar distances, positions and velocities. In contrast, earlier estimates of the Milky Way's rotation curve relied on ground based telescopes or earlier space mission observations, resulting in less precise distance and velocity measurements, and large uncertainties. This could explain the difference between the sharply declining Milky Way rotation curve and the mainly flat rotation curves of external galaxies.

The Milky Way galaxy Gaia DR3 rotation curve is fitted by Newtonian gravity, as shown in Figure \ref{fig:main} and Figure \ref{fig:milkyway}. It is possible that if the CGM plasma-gas continues to be confirmed experimentally~\cite{Nicastro2}, then the additional baryon mass required to account for the estimated Newtonian total Milky Way mass,
$M\sim 2\times 10^{11}M_{\odot}$, could be attributed to the CGM plasma-gas and not dark matter. This follows from a visible baryon mass estimate of $M_b\sim(0.6-1.0)\times10^{11}$~\cite{Nicastro1} and the estimated mass of the CGM plasma-gas, $M_{\rm CGM}\simeq (1-1.7)\times10^{11}M_\odot(Z/0.3Z_\odot)^{-1}$~\cite{Nicastro2}.

If further Gaia data releases confirm the new estimated mass and the Keplerian decline of the Milky Way rotation curve, and the excess baryon mass is in the form of a hot baryon gas enveloping the Milky Way, then there is an issue with our current understanding of the dynamics and evolution of galaxies, and the role of dark matter in the $\Lambda$CDM model.

In future work, we will investigate fitting the Milky Way rotation curve on the basis of a mass distribution formed from a spherically symmetric bulge and an axisymmetric thin disk. 

\section*{Acknowledgments}

 We thank Martin Green for helpful discussions. Research at the Perimeter Institute for Theoretical Physics is supported by the Government of Canada through industry Canada and by the Province of Ontario through the Ministry of Research and Innovation (MRI).

\end{document}